\documentclass[11pt]{article}
\usepackage{mathptmx}
\usepackage{sprinconf}
\usepackage{graphicx}
\usepackage{bm}
\usepackage{amsmath}
\newcommand{\half}{ \frac{1}{2} }
\newcommand{\dotX}{\dot{X}}

\begin{document}
\renewcommand{\thefootnote}{\fnsymbol{footnote}}
\title{Spectrum-Generating Algebra for Charged Strings\footnote{Talk presented at 26th International Colloquium on Group Theoretical Methodes\\
\hspace*{1.4pc} in Physics, CUNY, New York, June 2006.}}
\renewcommand{\thefootnote}{\arabic{footnote}}
\author{A. Kokado\dag\footnote[1]{kokado@kobe-kiu.ac.jp},
G. Konisi\ddag\footnote{konisi@wombat.zaq.ne.jp}
and T. Saito\ddag\footnote{tsaito@k7.dion.ne.jp}
}
\affil{\dag\ Kobe International University, Kobe 658-0032, Japan
}
\vspace{-36pt}
\affil{\ddag\ Department of Physics, Kwansei Gakuin University,
Sanda 669-1337, Japan
}
\beginabstract
When an open string ends with charges on a D2-brane,
which involves constant background magnetic field
perpendicular to the brane,
we construct the spectrum-generating algebra for this charged string,
which assures that our system is ghost-free under some conditions.
The application to the Hall effect for charged strings is also shortly 
remarked.
\endabstract
%
\section{Introduction\label{intro}}
We consider the Hall system for charged strings on the D-2 brane.
The charged open string is defined as $X^\mu(\tau, \sigma)$,~
$0 \le \sigma \le \pi$\,, which ends on the D2-brane
with different charges $q_0$ at $\sigma =0$
and $q_\pi $ at $\sigma =\pi$,
whereas $q_0 +q_\pi =0$ for the neutral string.
The D2-brane involves a constant magnetic field $B$
perpendicular to the brane.

Contrary to the neutral string,
the quantization of the charged string is not so well known%
\cite{AbouelsaoodEtAL},\cite{chu2000},\cite{KokadoEtAL2000}.
Here we meet a problem of how to work the no-ghost theorem
in our system.
For the neutral string with a background magnetic field
the spectrum-generating algebra (SGA) is well known
to assure the ghost-free for string states
if the space-time dimension is $d=26$ and the Regge intercept
is $\alpha (0)=1$.
For the charged string with a background field,
however, it is not clear whether the conventional SGA works
well or not, because of existence of normal modes
involving the cyclotron frequency.
We then consider the modification of SGA, and find that our system
is ghost-free if $d=26$ and $\alpha(0) =1-|\omega |/2+\omega ^2/2$,
where $\omega $ is the cyclotron frequency of the charged string.
\section{Quantization of the charged string\label{sec:1}}
The interaction Lagrangian of our system is given by
\begin{equation}
   L_I
  = q_\sigma \, \dotX_\mu(\tau, \sigma)\,
    A^\mu\big( X(\tau, \sigma) \big)~\Big\vert_{\sigma=0}
  + q_\sigma \, \dotX_\mu(\tau, \sigma)\,
    A^\mu\big( X(\tau, \sigma) \big)~\Big\vert_{\sigma=\pi }~.
\label{eq:def-L_int}
\end{equation}
We have chosen a gauge such as
$A^\mu = - (1/2)\, F^{\mu}{}_{\nu}~X^\nu$
with $F^{\mu}{}_{\nu}$, the constant electromagnetic field strength.
We will concentrate on a $3\times 3$ block of $F^{\mu}{}_{\nu}$,
\begin{align}
  & F^{\mu}{}_{\nu}
  = \begin{pmatrix} 0 &  E & 0 \\ E & 0 & B \\ 0 & - B & 0
    \end{pmatrix}~.
\label{eq:def-F}
\end{align}

Introducing a time-independent function $\rho(\sigma)$ such that
$\rho(0) = q_0,~\rho(\pi) = -q_\pi$,
the interaction Lagrangian can be rewritten as
\begin{align}
   L_I
  &= \half~\Big[~\rho(\sigma)\, \dotX_\mu(\tau, \sigma)\,
  F^{\mu}{}_{\nu}\, X^\nu(\tau, \sigma)~\Big]^{\sigma=\pi}_{\sigma=0}
  = \half \int^\pi_0 d\sigma~\partial_\sigma \Big(
     \rho\, \dotX_\mu\, F^{\mu}{}_{\nu}\, X^\nu \Big)
\nonumber \\
  &= \half \int^\pi_0 d\sigma~\Big[~
     \rho'\, \dotX_\mu\, F^{\mu}{}_{\nu}\, X^\nu
  + 2\, \rho\, \dotX_\mu\, F^{\mu}{}_{\nu}\, \big( X^\nu \big)'~\Big]~,
\label{eq:def-L_int-II}
\end{align}
where the total time derivative $\partial_\tau\big( \rho X' F X \big)$
has been discarded.
This theory does not depend on the functional form of $\rho(\sigma)$
for $0 < \sigma < \pi$,
because the action based on Eq.(\ref{eq:def-L_int-II})
is invariant under a variation with respect to $\rho(\sigma)$.

The total Lagrangian is given as
\begin{align}
  & L
  = \half \int^\pi_0 d\sigma~\left( \dotX_\mu\, \dotX^\mu
     - {X_\mu}' {X^\mu}' \right)
  + \int^\pi_0 d\sigma~\dotX_\mu F^{\mu}{}_{\nu} \left(
     \frac{ \rho'\, X^\nu }{2} + \rho\,  {X^\nu}' \right)~.
\label{eq:def-L}
\end{align}
First of all we diagonalize $F$ by a matrix $S$ as

\begin{align}
  & \left( S^{-1} F S \right)^\mu{}_\nu
  = \begin{pmatrix} 0 &     0 & 0 \\
                    0 & -i\, K & 0 \\
                    0 &     0 &  i\, K \end{pmatrix}~,
\label{eq:S_inv-F-S}
\end{align}
where $K = \sqrt{ B^2 - E^2 }$~(we assume $0 < E < B$).
Then we define relevant fields by

\begin{align}
  & S^{-1} \begin{pmatrix} X^0 \\ X^1 \\ X^2 \end{pmatrix}
  \equiv  \begin{pmatrix} Z \\ X^{(+)} \\ X^{(-)} \end{pmatrix}~.
\label{eq:def-ZX}
\end{align}

The boundary conditions for these variables become
\begin{align}
  & \big( X'^{(\pm )} \pm i\, \rho K \dot X^{(\pm )} \big)
    \Big\vert_{\sigma=0, \pi} = 0~,
& & Z'(\tau ,\sigma )\, \Big\vert_{\sigma=0, \pi} = 0~,
\nonumber \\
  & X^{a}(\tau ,\sigma ) \Big\vert_{\sigma=0, \pi}=c^a(const.),
  \quad a=3,\cdots ,d-1~.
\label{eq:XpmX'}
\end{align}
The Dirac quantization for this constrained system
has been carried out in Ref.\cite{KokadoEtAL2000}.
The result is summarized as follows:
Mode expansions for $X^{(\pm)}(\tau ,\sigma )$
satisfying the boundary conditions are
\begin{align}
  & X^{(\pm)}(\tau, \sigma)
  =  \sum_n\, \frac{ \cos\big[~(n \pm \omega) \sigma
                               \mp \pi \omega_0~\big] }
                   {-i (n \pm \omega) }~e^{ -i (n \pm \omega) \tau}~
     \alpha^{(\pm)}_n + b^{(\pm)}~,
\label{eq:X-mode_decomposition}
\end{align}
where $\tan \pi \omega _0 \equiv \rho (0)K,
\tan \pi \omega _\pi \equiv \rho (\pi )K, \omega
= \omega _0-\omega _\pi$ (we assume $\omega >0$).
On the other hand, $Z(\tau ,\sigma )$ does not couple with any field,
so it is a free string, as well as other components,
$X^3,\cdot \cdot \cdot ,X^{d-1}$.
The commutation relations for mode operators for $X^{(\pm)}$ are
\begin{align}
  & \big[\, \alpha^{(\pm)}_m\,,~\alpha^{(\mp)}_n~\big]
  = ( m \pm \omega  )~\delta_{m+n, 0}~,
& & \big[\, b^{(\pm)}\,,~b^{(\mp)}~\big]
  = - \frac{\cos \pi \omega _0 \cos \pi \omega _\pi }
           {\sin \pi \omega }~,
\label{eq:CCR_alpha}
\end{align}
whereas for other free components, $Z,X^3,\cdots ,X^{d-2}$,
we have usual ones.
From Eqs. \ref{eq:CCR_alpha}
we find the noncommutativity of $X^{(\pm)}(\tau ,\sigma )$
at $\sigma =0,\pi$.

The Virasoro operator is defined as
\begin{align}
  & L_n
  = \frac{1}{4} \int^{\pi}_{-\pi} d\sigma~e^{\pm i n \sigma}~
      : \big( \dotX \pm X' \big)^2 :~
  = \half~\sum_{l}: \alpha_l \cdot \alpha_{n-l} :~,
\label{eq:def-Virasoro_op}
\end{align}
where $A\cdot B=-A^0B^0 + A^{(+)}B^{(-)} + A^{(-)}B^{(+)} +A^3B^3
+\cdot \cdot \cdot +A^{d-1}B^{d-1} $,
and  satisfies the Virasoro algebra%
\begin{align}
  & \big[\, L_m\,,~L_n~\big]
  = (m-n)~L_{m+n} + m\, \left\{\, \frac{d}{12}\, (m^2-1)
  - \omega ^2 + \omega \, \right\}\, \delta _{m+n,0}~.
\label{eq:CCR_alphamLn}
\end{align}
%
\section{Spectrum-generating algebra for charged string \label{sec:2}}
We consider the string coordinates,
$Z, X^{(+)}, X^{(-)}, X^{3}, \cdots, X^{d-2}, X^{d-1}$.
The light-cone variables are now
$X^{\pm}\equiv (Z \pm X^{d-1})/\sqrt{2}$
(which should be distinguished from $X^{(\pm)}$), and
\begin{align}
  X^{\pm}(\tau )
  = \sum_{n \neq 0}\,\frac{ e^{-in\tau } }{n}\,
    \alpha^{\pm}_n + x^{\pm} + \tau p^{\pm}~,
 \ \ P^{\pm}(\tau )
  = \dot{X}^{\pm}(\tau )
  =\sum_{n}\,e^{-in\tau } \alpha^{\pm}_n~.
\label{eq:P_mode_exp}
\end{align}
For the $(\pm)$ components, we adopt
\begin{align}
  P^{(\pm)}(\tau )
  = \sum_{n}\,e^{-i(n\pm \omega )\tau } \alpha^{(\pm)}_n~,
\label{eq:P_mode_expII}
\end{align}
which satisfy
\begin{align}
  & \big[\, P^{(\pm)}(\tau )\,,~P^{(\mp)}(\tau' )\big]
  = \sum_{n}\,(n \pm \omega ) e^{-i(n\pm \omega )(\tau - \tau')}
  = i\, \partial _\tau \delta_{\pm \omega } (\tau - \tau')~.
\label{eq:CC_PP}
\end{align}
Here $\delta _{\omega }(s)$ is defined
as $\delta _{\omega }(s)=\exp(-i\omega s)\delta(s)$
with 2$\pi$-period delta function $\delta(s)$,
and has a property, $\delta _{\omega }(s+2\pi )
=\exp(-2\pi i\omega)\delta_{\omega }(s)$.

We define the SGA operators for the $(\pm)$ components as
\begin{align}
 & A^{(\pm )}_n
 = \frac{1}{2 \pi}\, \int^\pi_{-\pi} d\tau~
    P^{(\pm )}(\tau )V(\tau )^{n\pm \omega }~,
\label{eq:Apm=def}
\end{align}
where $V(\tau )=\exp[i X^-(\tau )]$.
For the sake of $p^{-}=1$, $V(\tau )$ carries
a factor $e^{i\tau }$ and hence $V(\tau )^{n \pm \omega}$
a factor $e^{i(n\pm \omega )\tau }$.
The non-periodic factor $e^{i\omega \tau }$
is canceled by the similar factor in  $P^{(\pm)}(\tau)$.
In the following calculations this cancellation
always occurs among $\delta _{\pm \omega }, V(\tau )^{n \pm \omega}$
and $P^{(\pm)}$.

We can then show that the modified SGA for the $(\pm)$-components
is given by
\begin{align}
  & \big[\, A^{(\pm)}_m \,,~A^{(\mp)}_n ~\big]
  =(m \pm \omega )~\delta _{m+n,0}~,
\nonumber \\
  & \big[\, A^{(\pm )}_m\,,~K_n~\big]
  = (m\pm \omega )~A^{(\pm)}_{m+n}~,
\label{eq:CCR_AmKn-III} \\
  & \big[\, K_m\,,~K_n~\big]
  = (m - n )~K_{m+n} + 2m^3 \delta _{m+n,0}~,
\nonumber
\end{align}
where
\begin{align}
  & K_n
  = - \frac{1}{2 \pi}\, \int^\pi_{-\pi} d\tau~: \left\{~P^{+}(\tau )
  + \frac{1}{2}\, n^2P^{-}(\tau )\log P^{-}(\tau )\right\}
     V(\tau )^n~:~,
\label{eq:K-def}
\end{align}
$P^{\pm}, X^{\pm}$ being light-cone variables.
We can also show that $A^{(\pm)}_n, K_n$ are commutable
with the Virasoro operator $L_n$ for the charged string.

These are to be compared with the commutation relations
\begin{align}
  & \big[\, \alpha ^{(\pm)}_m \,,~\alpha ^{(\mp)}_n ~\big]
  = (m \pm \omega )\, \delta _{m+n,0}~,
\nonumber \\
  & \big[\, \alpha ^{(\pm)}_m\,,~L_n^T~\big]
  = (m \pm \omega )~\alpha ^{(\pm)}_{m+n}~,
\label{eq:CCR_alphamLnT} \\
  & \big[\, L_m^T\,,~L_n^T~\big]
  = (m-n)~L_{m+n}^T
  + m\, \left\{\, \frac{d-2}{12}\, (m^2-1)
  - (\omega ^2-\omega -2a)\, \right\}\, \delta _{m+n,0}~,
\nonumber
\end{align}
where
\begin{align}
   L_{n}^{T}
  &\equiv  -a\, \delta _{n,0}
\nonumber \\
  &+ \frac{1}{2}\, \sum_{l}:[\, \alpha ^{(+)}_{n-l}\, \alpha ^{(-)}_{l}
  + \alpha ^{(-)}_{n-l}\, \alpha^{(+)}_{l}
  + \alpha ^3_{n-l}\, \alpha ^3_{l}
  + \cdots + \alpha ^{d-2}_{n-l}\, \alpha ^{d-2}_{l}]:~.
\label{eq:L-exp-II}
\end{align}
Comparing Eqs.(\ref{eq:CCR_AmKn-III}) with (\ref{eq:CCR_alphamLnT}),
the isomorphism
\begin{align}
  & A_n^{(\pm)} \sim \alpha _n^{(\pm)}~,
& & K_n \sim L_n^{T}~,
\label{eq:isomorphism2}
\end{align}
is completed if
\begin{align}
  & d=26~,
& & a \equiv \alpha (0)
  =1-\omega/2 + \omega ^2/2~.
\label{eq:d_alpha-condition}
\end{align}
The other transverse components obey the same algebra
as in the free case, and we have the same isomorphism
as Eqs.(\ref{eq:isomorphism2}).
Once we have SGA, then we can conclude that our system
is ghost-free provided the conditions (\ref{eq:d_alpha-condition})
are satisfied.

When $F^\mu _{\ \nu }$ has many blocks, the terms $\omega ^2 - \omega $ appearing 
in the last equation of Eqs.(\ref{eq:CCR_alphamLnT}) should be replaced 
by sum $\sum_i ({\omega _i}^2 - \omega_i )$, where $\omega _i$ is the cyclotron 
frequency of the $i$-th block. The same is true for Eq.(\ref{eq:d_alpha-condition})
\section{Concluding remarks \label{sec:3}}
We have constructed SGA for charged strings
with a constant background magnetic fields.
We conclude that our system is ghost-free
if the space-time dimension is $d=26$ and
the Regge intercept is $\alpha (0)=1-\omega /2 + \omega ^2/2$,
where $\omega $ is the cyclotron frequency of the charged string.
If we take a limit $\omega \rightarrow 0$,
then the above results are all reduced to
the usual results of neutral strings. 
See also Ref.\cite{KokadoEtAL2000}. 

Finally let us remark about the Hall conductivity
for the charged strings,
which end on the D2-brane with a constant background magnetic field
$B$ perpendicular to the brane.
The result is given by $\sigma _H=-(q_0+q_\pi )n/B$,
$n$ being the charged string density, which coincides completely
with that of the ordinary two-dimensional electron system%
\cite{KokadoEtAL2006}.

\end{document}